\documentclass[onecolumn,prb]{revtex4}
\usepackage{graphicx}
\usepackage{dcolumn}
\usepackage{bm}
\usepackage{amsmath}

\begin{document}

\preprint{}
\title{Influence of magnetic impurities on charge transport in
diffusive-normal-metal / superconductor junctions}
\author{T. Yokoyama$^1$, Y. Tanaka$^1$, A. A. Golubov$^2$, J. Inoue$^1$ and
Y. Asano$^3$}
\affiliation{$^1$Department of Applied Physics, Nagoya University, Nagoya, 464-8603, Japan%
\\
CREST, Japan Science and Technology Corporation (JST) 
Nagoya, 464-8603, Japan \\
$^2$ Faculty of Science and Technology, University of Twente, 7500 AE,
Enschede, The Netherlands\\
$^3$Department of Applied Physics, Hokkaido University,Sapporo, 060-8628,
Japan}
\date{\today}

\begin{abstract}
Charge transport in the diffusive normal metal (DN) / insulator / $s$- and $%
d $-wave superconductor junctions is studied in the presence of magnetic
impurities in DN in the framework of the quasiclassical Usadel equations
with the generalized boundary conditions. The cases of $s$- and $d$-wave
superconducting electrodes are considered. The junction conductance is
calculated as a function of a bias voltage for various parameters of the DN
metal: resistivity, Thouless energy, the magnetic impurity scattering rate and the
transparency of the insulating barrier between DN and a superconductor. It
is shown that the proximity effect is suppressed by magnetic impurity
scattering in DN for any value of the barrier transparency. In
low-transparent $s$-wave junctions this leads to the suppression of the
normalized zero-bias conductance. In contrast to that, in high transparent
junctions zero-bias conductance is enhanced by magnetic impurity scattering. The
physical origin of this effect is discussed. For the $d$-wave junctions, the
dependence on the misorientation angle $\alpha$ between the interface normal
and the crystal axis of a superconductor is studied. The zero-bias
conductance peak is suppressed by the magnetic impurity scattering only for low
transparent junctions with $\alpha \sim 0$. In other cases the conductance
of the $d$-wave junctions does not depend on the magnetic impurity scattering due to
strong suppression of the proximity effect by the midgap Andreev resonant
states.
\end{abstract}

\maketitle



%

%




\section{Introduction}

Nowadays, thanks to the nanofabrication technique, detailed experimental
studies of the electron coherence in mesoscopic superconducting systems
become possible, where the Andreev reflection \cite{Andreev,BTK,Zaitsev}
plays an important role in the low energy transport.
In diffusive normal metal / superconductor (DN/S) junctions, the phase
coherence between incoming electrons and Andreev reflected holes persists in
DN at a mesoscopic length scale and results in strong interference effects
on the probability of Andreev reflection \cite{Hekking}.

One of the remarkable experimental manifestations of the coherent Andreev
reflection is the zero bias conductance peak (ZBCP) in DN/S junctions \cite%
{Giazotto,Klapwijk,Kastalsky,Nguyen,Wees,Nitta,Bakker,Xiong,Magnee,Kutch,Poirier}%
. The physics of ZBCP was extensively studied theoretically using scattering
matrix approach \cite{Beenakker1,Lambert,Takane,Beenakker2,reflec,Lesovik}
and the quasiclassical Green's function technique \cite%
{Volkov,Nazarov1,Yip,Yip2,Stoof,Reentrance,Golubov,Takayanagi,Bezuglyi,Seviour,Belzig}%
.
Volkov, Zaitsev and Klapwijk (VZK) \cite{Volkov} explained the origin of the
ZBCP in DN/S junctions in the framework of the quasiclassical theory by
solving the Usadel equations \cite{Usadel} with the Kupriyanov and Lukichev
(KL) boundary condition for the Keldysh-Nambu Green's function \cite{KL}.
According to the VZK theory the ZBCP is due to the enhancement of the pair
amplitude in DN by the proximity effect. The influence of the magnetic
impurity scattering on the bias voltage dependent conductance was also
studied within this approach \cite{Volkov,Yip2,Belzig1}.

%
%
Recently the VZK theory for $s$-wave superconductors was extended by Tanaka
et al. \cite{TGK} using more general boundary conditions provided by the
circuit theory of Nazarov \cite{Nazarov2}. These boundary conditions treat
an interface as an arbitrary connector between diffusive metals. The
connector is characterized by a set of transmission coefficients ranging
from a ballistic point contact to a tunnel junction. 
The boundary conditions coincide with the KL conditions when a
connector is diffusive or transmission coefficients are low, while
the BTK theory \cite{BTK} is reproduced in the ballistic regime. The
extended VZK theory \cite{TGK,PRB2004} revealed a number of new
features like a $U$ -shaped gap like structure and a crossover from
a zero bias conductance peak (ZBCP) to a zero bias conductance dip
(ZBCD). These phenomena are relevant for the actual junctions in
which the barrier transparency is not necessarily small. However,
the influence of the magnetic impurity scattering in DN on the
charge transport was not studied in this regime.

The generalized VZK theory was recently applied also to unconventional
superconducting junctions \cite{Nazarov3,PRB2004}. The formation of the
midgap Andreev resonant states (MARS) at the interface of unconventional
superconductors \cite{Buch,TK95,Kashi00,Experiments} is naturally taken into
account in this approach \cite{Nazarov3,PRB2004}. It was demonstrated that
the formation of MARS in DN/$d$-wave superconductor(DN/$d$) junctions strongly competes with the proximity
effect. Remarkable recent advances in experiments on tunneling in high $T_{C}
$ cuprates \cite{Hiromi} stimulate an interest to the problem of an
influence of the magnetic impurity scattering on a charge transport in DN/$d$
junctions.

In the present paper the generalized VZK theory is applied to the
study of an influence of the magnetic impurity scattering in the
DN on the conductance in DN/S where S is either $s$- or
$d$-wave superconductor. The parameters of the problem are the
height of the insulating barrier at the DN/S interface, the
resistance $R_{d}$, the magnetic impurity scattering rate $\gamma$, the Thouless
energy $E_{Th}$ in DN and the angle  $\alpha $ between the normal to
the interface and the crystal axis of $d$-wave superconductors. We
shall focus on the dependence of the normalized conductance $\sigma
_{T}(eV)=\sigma _{S}(eV)/\sigma _{N}(eV)$, on the bias voltage $V$, where $%
\sigma _{S(N)}(eV)$ are the conductances in the superconducting (normal )
state.
The organization of the paper is as follows. In section II the detailed
derivation of the expression for the normalized conductance is provided. In
sections III the results of calculations of $\sigma _{T}(eV)$ are presented
for $s$- and $d$-wave junctions separately and physical explanation of the
results is given. In section IV the summary of the obtained results and the
conclusions are presented. In this paper we restrict ourselves to the
low-temperature regime $T<<T_{c}$ and put $k_{B}=\hbar =1$.

\section{Formulation}

In this section we introduce the model and the formalism. We consider a
junction consisting of normal and superconducting reservoirs connected by a
quasi-one-dimensional diffusive conductor (DN) with a length $L$ much larger
than the mean free path. This structure was considered in Ref.~\cite%
{TGK,PRB2004}, while in the present paper the scattering on magnetic
impurities in DN is taken into account. Similar to Ref.~\cite{TGK,PRB2004},
we assume that the interface between the DN conductor and the S electrode at
$x=L$ has a resistance $R_{b}$ while the DN/N interface at $x=0$ has zero
resistance and we apply the generalized boundary conditions of Ref.~\cite%
{Nazarov2} to treat the interface between DN and S.

\begin{figure}[tbh]
\begin{center}
\scalebox{0.4}{
\includegraphics[width=20.0cm,clip]{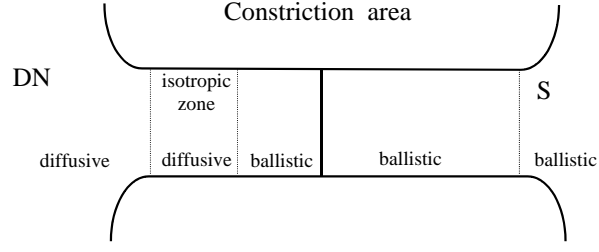}}
\end{center}
\caption{ Schematic illustration of the model}
\label{f0}
\end{figure}

We model the insulating barrier between DN and S by the delta function $%
U(x)=H\delta (x-L)$, which provides the transparency of the junction $%
T_{m}=4\cos ^{2}\phi /(4\cos ^{2}\phi +Z^{2})$, where $Z=2H/v_{F}$ is a
dimensionless constant, $\phi $ is the injection angle measured from the
interface normal to the junction and $v_{F}$ is Fermi velocity. The
interface resistance $R_{b}$ is given by
\begin{equation*}
R_{b}=R_{0}\frac{2}{\int_{-\pi /2}^{\pi /2}d\phi T_{m}\cos \phi },
\end{equation*}%
where $R_{0}$ is Sharvin resistance $R_{0}^{-1}=e^{2}k_{F}^{2}S_{c}/4\pi ^{2}
$, $k_{F}$ is the Fermi wave-vector and $S_{c}$ is the constriction area
(see Fig. \ref{f0}). Note that the area $S_{c}$ is in general not equal to
the cross-section area $S_{d}$ of the normal conductor, therefore $%
S_{c}/S_{d}$ is independent parameter of our theory. This allows to vary $%
R_{d}/R_{b}$ independently of $T_{m}$. In real physical situation, the
assumption $S_{c}<S_{d}$ means that only a part of the actual flat DN/S
interface (having area $S_{c}$) is conducting, no matter is it a single
conducting region or a series of such regions. These conducting regions are
not constrictions in a standard sense - we don't assume the narrowing of the
total cross-section, but rather that only the part of the cross-section is
conducting.

We apply the quasiclassical Keldysh formalism in the following calculation
of the conductance. The definitions of 4 $\times $ 4 Green's functions in DN
and S, $\check{G}_{1}(x)$ and $\check{G}_{2}(x)$, and other notations can be
found in Ref.~\cite{TGK,PRB2004}.
The new feature in the present model is the spin-scattering term in the
static Usadel equation \cite{Usadel} for $\check{G}_{1}(x)$ in DN
\begin{equation}
D\frac{\partial }{\partial x}[\check{G}_{1}(x)\frac{\partial \check{G}_{1}(x)%
}{\partial x}]+i[\check{H}-i\check{\Sigma}_{spin},\check{G}_{1}(x)]=0,
\end{equation}%
where $D$ is the diffusion constant in DN, $\check{H}$ is given by
\begin{equation*}
\check{H}=\left(
\begin{array}{cc}
\hat{H}_{0} & 0 \\
0 & \hat{H}_{0}%
\end{array}%
\right) ,
\end{equation*}%
with $\hat{H}_{0}=\epsilon \hat{\tau}_{3}$, and
\begin{equation*}
\check{\Sigma}_{spin}=\frac{\gamma }{2}\hat{\tau}_{3}\check{G}_{1}(x)\hat{%
\tau}_{3}
\end{equation*}%
is the self-energy for magnetic impurity scattering with the scattering rate $\gamma $
in DN. Note that magnetic impurities take random alignments and we average
them in all directions, thus $\check{G}_{1}(x)$ in our calculation is a unit
matrix in the spin space. The Nazarov's generalized boundary condition for $%
\check{G}_{1}(x)$ at the DN/S interface has the same form as the one without
magnetic impurity scattering (see Ref.~\cite{TGK,PRB2004}).

In the actual calculation it is convenient to use the standard $\theta$%
-parametrization where $\theta (x)$ is a measure of the proximity effect in
DN and is determined by the following equation
\begin{equation}
D\frac{\partial ^{2}}{\partial x^{2}}\theta (x)+2i(\epsilon + i\gamma \cos
[\theta (x)])\sin [\theta (x)]=0,  \label{Usa1}
\end{equation}%
One can see that introduction of magnetic impurity scattering $\gamma$ leads to
modification of the effective coherence length in DN. In particular,
switching on $\gamma$ makes function $\theta (x)$ exponentially decaying at
zero energy, while $\theta (x)$ at $\gamma=0$ behaves linearly in DN. It
will be shown below that these modifications result in suppression of $%
\theta $ in DN, as expected due to pair-breaking nature of magnetic
scattering, which in turn leads to corresponding modifications of the subgap
conductance.

%
Finally, we obtain the following result for the electric current
\begin{equation}
I_{el}=\frac{1}{e}\int_{0}^{\infty }d\epsilon \frac{f_{t0}}{\frac{R_{b}}{%
<I_{b0}>}+\frac{R_{d}}{L}\int_{0}^{L}\frac{dx}{\cosh ^{2} \theta _{im}(x)}}.
\end{equation}%
Then the total resistance $R$ for $s$-wave junction at zero temperature is
given by
\begin{equation}
R=\frac{R_{b}}{<I_{b0}>}+\frac{R_{d}}{L}\int_{0}^{L}\frac{dx}{\cosh
^{2}\theta _{im}(x)}
\end{equation}
with
\begin{equation*}
I_{b0} = \frac{T_{m} \Lambda_{1} + 2(2-T_{m}) \Lambda_{2}} {2 \mid (2-T_{m})
+ T_{m}[g \cos\theta_{L} + f \sin\theta_{L} ] \mid^{2} },
\end{equation*}
\begin{equation*}
\Lambda_{1}=(1+\mid \cos\theta_{L} \mid^{2} + \mid \sin\theta_{L} \mid^{2})
(\mid g \mid^{2} + \mid f \mid^{2} +1)
\end{equation*}
\begin{equation}
+ 4\mathrm{Imag}[fg^{*}] \mathrm{Imag}[\cos \theta_{L} \sin\theta_{L}^{*} ],
\end{equation}

\begin{equation}
\Lambda_{2} =\mathrm{{Real} \{ g(\cos \theta_{L} + \cos \theta_{L}^{*}) +
f(\sin \theta_{L} + \sin \theta_{L}^{*}) \}},
\end{equation}

\begin{equation*}
g=\varepsilon /\sqrt{\varepsilon ^{2}-\Delta^{2}}, f=\Delta/\sqrt{
\Delta^{2}-\varepsilon ^{2}}.
\end{equation*}

For a $d$-wave junction, the function $I_{b0}$ is given by the following
expression
\begin{equation*}
I_{b0}= \frac{T_{n}}{2} \frac{C_{0}}{ \mid
(2-T_{n})(1+g_{+}g_{-}+f_{+}f_{-}) +T_{n}[\cos\theta_{L} (g_{+} +g_{-}) +
\sin\theta_{L}(f_{+} + f_{-})] \mid^{2}}
\end{equation*}
\begin{equation*}
C_{0} =T_{n}(1+\mid \cos\theta_{L} \mid^{2} + \mid \sin\theta_{L} \mid^{2})
[\mid g_{+} +g_{-} \mid^{2} + \mid f_{+} +f_{-} \mid^{2} + \mid 1+
f_{+}f_{-} + g_{+}g_{-} \mid^{2} + \mid f_{+}g_{-} - g_{+}f_{-} \mid^{2} ]
\end{equation*}
\begin{equation*}
+ 2(2-T_{n})\mathrm{Real} \{(1+g_{+}^{*}g_{-}^{*}+f_{+}^{*}f_{-}^{*})
[(\cos\theta_{L} + \cos \theta^{*}_{L})(g_{+} + g_{-}) + (\sin\theta_{L} +
\sin \theta^{*}_{L})(f_{+} + f_{-}) ] \}
\end{equation*}
\begin{equation*}
+ 4T_{n}\mathrm{Imag}(\cos\theta_{L} \sin\theta^{*}_{L}) \mathrm{Imag}%
[(f_{+}+f_{-})(g_{+}^{*} + g_{-}^{*})],
\end{equation*}

$g_{\pm}=\varepsilon /\sqrt{\varepsilon ^{2}-\Delta _{\pm}^{2}}$, $
f_{\pm}=\Delta _{\pm}/\sqrt{\Delta _{\pm}^{2}-\varepsilon ^{2}}$ and $%
\Delta_{\pm} = \Delta\cos2(\phi \mp \alpha)$. In the above $\alpha$, $%
\theta_{im}(x)$ and $\theta_{L}$ denote the angle between the normal to the
interface and the crystal axis of $d$-wave superconductors, the imaginary
part of $\theta(x)$ and $\theta(L_{-})$ respectively. The conductance in the
superconducting state $\sigma _{S}(eV)$ is simply related to $R$ by $\sigma
_{S}(eV)=1/R$.

It is important to note that in the present approach, according to the
circuit theory, $R_{d}/R_{b}$ can be varied independently of $T_{m}$, $i.e.$%
, independently of $Z$, since one can change the magnitude of the
constriction area $S_c$ independently. In other words, $R_{d}/R_{b}$ is no
more proportional to $T_{av}(L/l)$, where $T_{av}$ is the averaged
transmissivity of the barrier and $l$ is the mean free path in the diffusive
region. Based on this fact, we can choose $R_{d}/R_{b}$ and $Z$ as
independent parameters.

In the following section, we will discuss the normalized conductance $\sigma
_{T}(eV)=\sigma _{S}(eV)/\sigma _{N}(eV)$ where $\sigma _{N}(eV)$ is the
conductance in the normal state without magnetic impurity given by $\sigma
_{N}(eV)=\sigma _{N}=1/(R_{d}+R_{b})$.


\section{Results}

\subsection{Tunneling conductance for $s$-wave junctions}

\begin{figure}[htb]
\begin{center}
\scalebox{0.4}{
\includegraphics[width=20.0cm,clip]{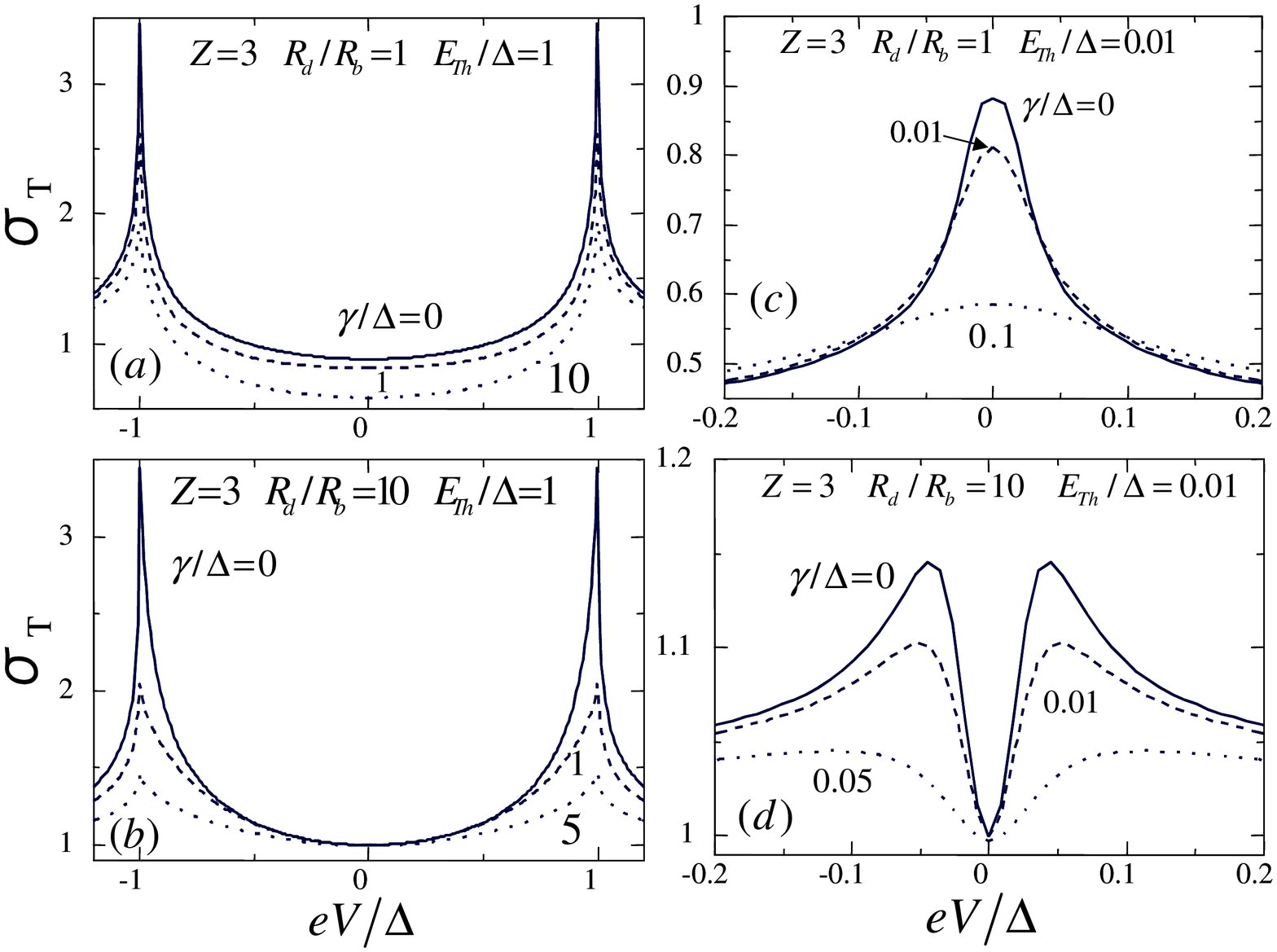}}
\end{center}
\par
\caption{ Normalized conductance for $Z=3$. }
\label{f2}
\end{figure}

In this section, we focus on the bias voltage dependent normalized
conductance $\sigma_{T}(eV)$ for various situations. Let us first focus on
the relatively low transparent junctions with $Z=3$ for various $%
\gamma/\Delta$(Fig. \ref{f2}). For $E_{Th}/\Delta=1$ and $R_d/R_b=1$, the $%
\sigma_{T}(eV)$ curves have a rounded bottom shape and the height of the
bottom value is reduced with an increase in $\gamma/\Delta$. The height of
the peak at $eV= \pm \Delta$ is reduced with an increase in $\gamma/\Delta$
(see Fig. \ref{f2}(a)). For $E_{Th}/\Delta=1$ and $R_d/R_b=10$, the $%
\sigma_{T}(eV)$ curves also have a rounded bottom structure which flattens
with an increase in $\gamma/\Delta$. Also the peak at $eV= \pm \Delta$ is
suppressed with the increase of $\gamma/\Delta$ (see Fig. \ref{f2}(b)). For
small Thouless energy $E_{Th}/\Delta=0.01$ and $R_d/R_b=1$, the conductance
has a prominent ZBCP with the width given by $E_{Th}$. As seen from Fig. \ref%
{f2}(c), the magnetic impurity scattering suppresses the peak height. With
the increase of the resistance ratio $R_d/R_b$, the ZBCP transforms into
ZBCD, as shown in Fig. \ref{f2}(d). The magnitude of ZBCD decreases with the
increase of $\gamma/\Delta$, and the height of the peaks around $eV/\Delta
\sim 0.04$ is also reduced (see Fig. \ref{f2} (d)). As seen from these
figures, the characteristic energy range of $\gamma$ which modifies the
magnitude of $\sigma _{T}(eV)$, is determined by $E_{Th}$, in agreement with
the previous study based on the KL boundary conditions \cite{Yip2}.

\begin{figure}[htb]
\begin{center}
\scalebox{0.4}{
\includegraphics[width=20.0cm,clip]{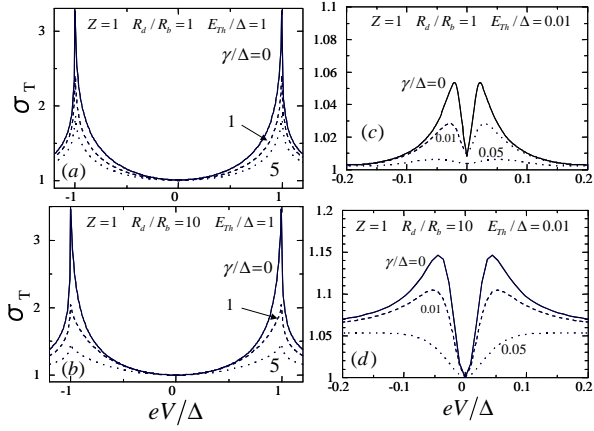}}
\end{center}
\par
\caption{ Normalized conductance for $Z=1$.}
\label{f3}
\end{figure}

In the case of an intermediate barrier strength $Z=1$(Fig. \ref{f3}) the
magnitude of $\sigma _{T}(eV)$ always exceeds unity. The resulting line
shapes of $\sigma_{T}(eV)$ for $E_{Th}/\Delta=1$ are quite similar to the
corresponding curves for $Z=3$ (see Figs. \ref{f3}(a) and \ref{f3}(b)). For $%
E_{Th}/\Delta=1$ and $R_d/R_b=1$, the zero-bias value $\sigma_{T}(0)$ is
independent of $\gamma/\Delta$ (see Fig \ref{f3}(a)), in contrast to the
corresponding case shown in Fig. \ref{f2}(a). Another important difference
from the case of large $Z$-factor is the absence of ZBCP for low Thouless
energy. It is seen that for $E_{Th}/\Delta=0.01$ a ZBCD occurs in both cases
of $R_d/R_b=1$ and $R_d/R_b=10$. This conductance dip and the finite voltage
peaks are fully suppressed with the increase of $\gamma/\Delta$ for $%
R_d/R_b=1$ (see Fig. \ref{f3}(c)). On the other hand, for $R_d/R_b=10$ only
the peaks around $eV/\Delta \sim 0.04$ are suppressed while the magnitude of
$\sigma(0)$ does not depend on $\gamma$, similar to the case $Z=3$ (see Fig. %
\ref{f3}(d)). The relevant scale of $\gamma$ is again given by the magnitude
of $E_{Th} $.

\begin{figure}[htb]
\begin{center}
\scalebox{0.4}{
\includegraphics[width=20.0cm,clip]{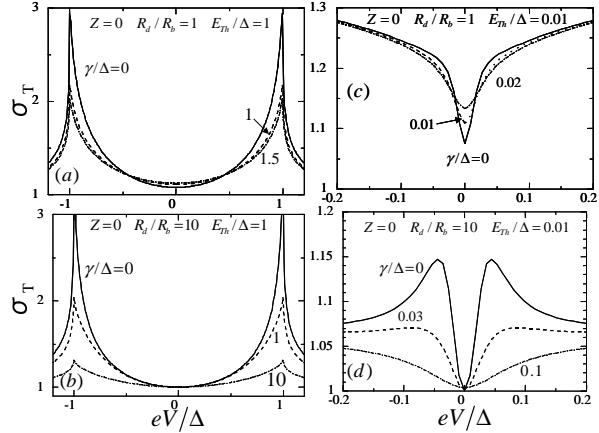}}
\end{center}
\par
\caption{ Normalized conductance for high transparent junctions with Z=0.}
\label{f4}
\end{figure}

%
For fully transparent case with $Z=0$(Fig. \ref{f4}), $\sigma _{T}(eV)$ also
always exceeds unity. The line shapes of $\sigma_{T}(eV)$ with $%
E_{Th}/\Delta=1$ are similar to the corresponding curves for $Z=3$ and $Z=1$%
(see Figs. \ref{f4}(a) and \ref{f4}(b)). For $E_{Th}/\Delta=1$ and $R_d/R_b=1
$, the magnitude of $\sigma_{T}(0)$ is enhanced by $\gamma/\Delta$ in
contrast to the corresponding cases shown in Figs. \ref{f2}(a) and \ref{f3}%
(a)(see Fig \ref{f4}(a)). For $E_{Th}/\Delta=0.01 $ and $R_d/R_b=1$, $%
\sigma_{T}(eV)$ have a ZBCD. The magnitude of $\sigma_{T}(0)$ is enhanced by
$\gamma/\Delta$ and the depth of the ZBCD decreases with the increase of $%
\gamma/\Delta$ (see Fig. \ref{f4}(c)). On the other hand, for $%
E_{Th}/\Delta=0.01$ and $R_d/R_b=10$, the magnitude of $\sigma(0)$ does not
depend on $\gamma$ while the finite bias peaks are suppressed similar to the
cases of $Z=3$ and $Z=1$ (see Fig. \ref{f4}(d)).

%

\begin{figure}[htb]
\begin{center}
\scalebox{0.45}{
\includegraphics[width=20.0cm,clip]{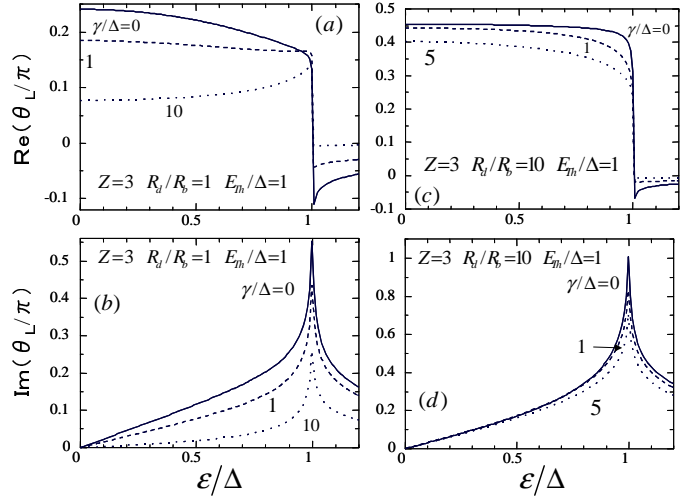}}
\end{center}
\caption{ Real (upper panels) and imaginary (lower panels) part of $\protect%
\theta _{L}$ for $Z=3$ and $E_{Th}/\Delta=1$. }
\label{f5}
\end{figure}

\begin{figure}[htb]
\begin{center}
\scalebox{0.45}{
\includegraphics[width=20.0cm,clip]{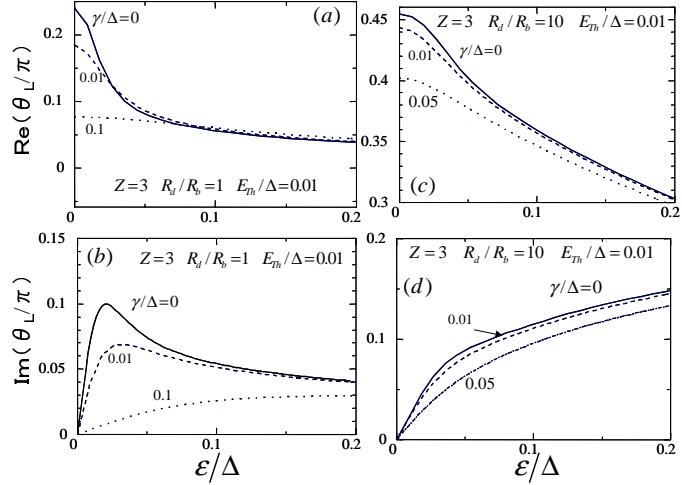}}
\end{center}
\caption{ Real (upper panels) and imaginary (lower panels) part of $\protect%
\theta _{L}$ for $Z=3$ and $E_{Th}/\Delta=0.01$. }
\label{f6}
\end{figure}

In order to understand the above wide variety of line shapes and their
relation to the proximity effect, we shall discuss the behavior of function $%
\theta_{L} $ which is the measure of the proximity effect at the DN/S
interface and determines the normalized local density of states by $%
Re\cos\theta(x)$. At $\epsilon =0$ $\theta _{L}$ is always a real number
even for non zero $\gamma$. First, we study the case of $Z=3$ and $%
E_{Th}/\Delta=1$ (Fig. \ref{f5}) where the same values of $\gamma/\Delta$
and $R_{d}/R_{b}$ are chosen as in Fig. \ref{f2}. The real part of $\theta
_{L}$ has a step function like structure and it is always positive for $%
\epsilon \le \Delta $ and negative otherwise. The absolute value of the real
part of $\theta _{L}$ decreases with an increase in $\gamma/\Delta$. At the
same time, the imaginary part of $\theta _{L}$ has a coherent peak, the
height of which is reduced with an increase in $\gamma/\Delta$. For the case
of $Z=3$ and $E_{Th}/\Delta=0.01$ (Fig. \ref{f6}) where the same values of $%
\gamma/\Delta$ are chosen as in Fig. \ref{f2}, the real part of $\theta _{L}$
has a ZBCP with the width given by $E_{Th}$. The imaginary part of $\theta
_{L}$ has a ZBCD for $R_d/R_b=1$. Both the amplitudes of the real and
imaginary part of $\theta _{L}$ are reduced with the increase of $%
\gamma/\Delta$ only around zero energy in the interval of the order of $%
E_{Th}$. 

Next we consider the case of $Z=0$ with $E_{Th}/\Delta=1$ (Fig. 7) and $%
E_{Th}/\Delta=0.01$ (Fig. 8) where the same values of $\gamma/\Delta$ are
chosen as in Fig. 4. The line shapes of both $\mathrm{Re}(\theta_{L})$ and $%
\mathrm{Im}(\theta_{L})$ are similar to those in Figs. \ref{f5} and \ref{f6}%
. There is no clear qualitative difference between the energy dependencies
of $\mathrm{\ Real[Imag]}(\theta _{L})$ for $Z=0$ and those for $Z=3$. For
all cases, the magnitude of $\theta_{L}$ is reduced with the increase of $%
\gamma$ and then the proximity effect is suppressed by the magnetic impurity
scattering within the energy range determined by $E_{Th}$. In almost all
cases, the magnitude of $\sigma_{T}(eV)$ is reduced with the decrease of $%
\theta_{L}$. Only for high transparent case with not so large $R_{d}/R_{b}$,
the decrease of the magnitude of $\theta_{L}$, $i.e.$, the reduction of the
proximity effect, can enhance the magnitude of $\sigma_{T}(eV)$.

\begin{figure}[htb]
\begin{center}
\scalebox{0.45}{
\includegraphics[width=20.0cm,clip]{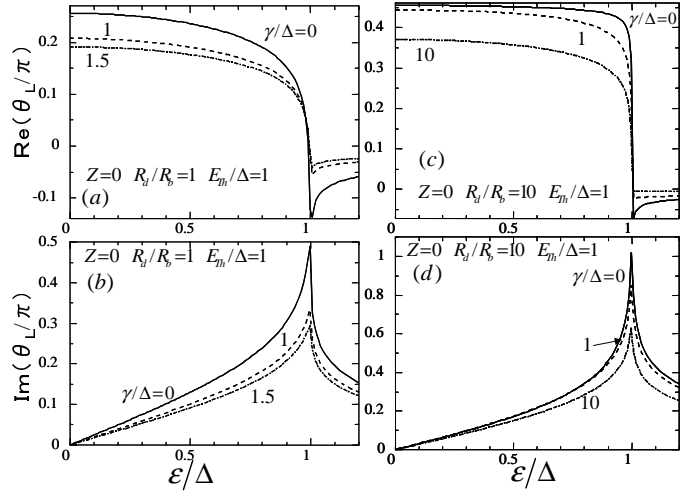}}
\end{center}
\caption{ Real (upper panels) and imaginary (lower panels) part of $\protect%
\theta _{L}$ for high transparent junctions with $Z=0$ and $E_{Th}/\Delta=1 $%
. }
\label{f7}
\end{figure}

\begin{figure}[htb]
\begin{center}
\scalebox{0.45}{
\includegraphics[width=20.0cm,clip]{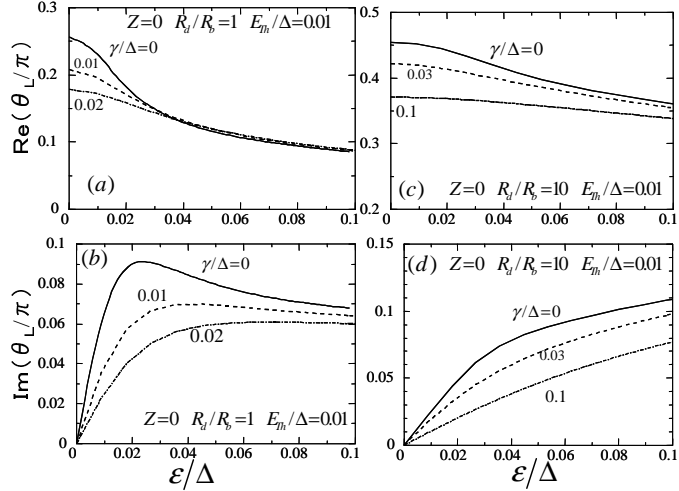}} \label{f8}
\end{center}
\caption{ Real (upper panels) and imaginary (lower panels) part of $\protect%
\theta _{L}$ for high transparent junctions with $Z=0$ and $%
E_{Th}/\Delta=0.01 $. }
\end{figure}
In the following, we explain the wide variety of the line shapes of $\sigma
_{T}(eV)$. We consider $Z=0$ and $E_{Th}/\Delta =1$ case, where $\theta _{L}$
has a weak energy dependence around zero voltage. For the fully transparent
case with $ T_{m}=1$ , i.e., $Z=0$, $\sigma _{T}(0)$ can be given by
\begin{equation}
\sigma _{T}(0)=\frac{1+R_{d}/R_{b}}{1/<I_{b0}>+R_{d}/R_{b}}
\end{equation}
with
\begin{equation}
<I_{b0}>=\frac{2}{1+\sin \theta _{L}}.
\end{equation}

From this equation we find that the magnitude of $\sigma _{T}(0)$ gets close
to unity under the strong proximity effect, i.e., when the magnitude of $%
R_{d}/R_{b}$ is large. 
As shown in Figs. \ref{f7}(a) and \ref{f7}(b), the magnitude of $\theta _{L}$
at $\epsilon =0$ is lowered with an increase in $\gamma /\Delta $ for $%
R_{d}/R_{b}=1$. Then , according to Eqs. (7) and (8), the resulting $\sigma
_{T}(eV)$ around $eV\sim 0$ is slightly enhanced as shown in Fig. \ref{f4}%
(a). For $R_{d}/R_{b}=10$, the magnitude of $R_{d}/R_{b}$ is much larger
than the magnitude of $1/<I_{b0}>$. Then the $\gamma $ dependence of $\sigma
_{T}(0)$ becomes negligible as shown in Fig. \ref{f4}(b). In order to
understand the case of $Z=0$ and the small magnitude of $E_{Th}/\Delta $, we
decompose $R$ into $R_{1}$ and $R_{2}$ following the previous work\cite{TGK}%
, where $R_{1}$ and $R_{2}$ are defined by 
\begin{equation*}
R_{1}=\frac{1}{L}\int_{0}^{L}\frac{dx}{\cosh ^{2}\theta _{im}(x)}
\end{equation*}%
and
\begin{equation*}
R_{2}=\frac{R_{b}}{R_{d}<I_{b0}>}.
\end{equation*}

Fig. 9 shows that $R_1$ has a minimum at a finite voltage which can result
in a ZBCD and that $R_2$ has a maximum for high transparent junctions. For a
large magnitude of $R_d/R_b$, the effect of $R_1$ is dominant, then the
normalized conductance always has a ZBCD (see Figs. 9(c), 9(d) and \ref{f4}%
(d)). Since $R_2$ has a maximum at zero voltage (Fig. 9(b)), the resulting $%
\sigma_{T}(eV)$ has a ZBCD as shown in Fig. \ref{f4}(c).

\begin{figure}[htb]
\begin{center}
\scalebox{0.45}{
\includegraphics[width=20.0cm,clip]{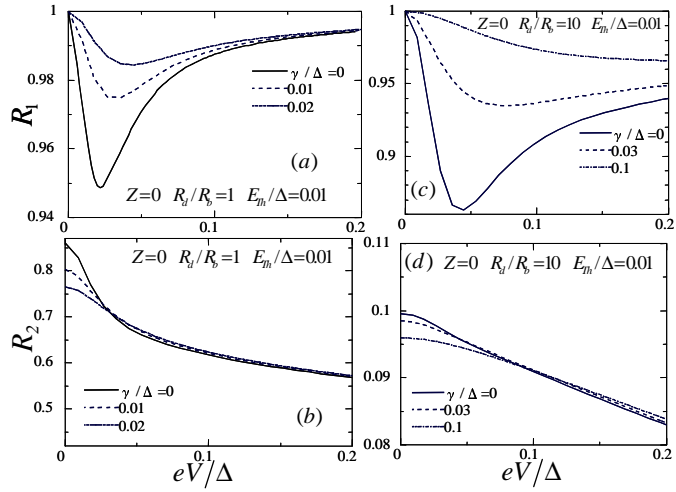}}
\end{center}
\par
\label{f9}
\caption{ Normalized resistance for $Z=0$ and $E_{Th}/\Delta=0.01$.}
\end{figure}
Next we focus on the zero voltage resistance $R/R_b$ as a function of $%
R_{d}/R_{b}$. For $Z=3$, $R/R_b$ has a reentrant behavior as a function of $%
R_{d}/R_{b}$ due to the so called reflectionless tunneling effect \cite%
{reflec} (see Fig. 10(a)). With the increase of $\gamma$, this effect is
smeared since the magnitude of $\theta_{L}$ is reduced as shown in Fig. 11.
In contrast, for $Z=0$, where $R/R_b$ increases monotonically as a function
of $R_{d}/R_{b}$, the $\gamma$ dependence of $R/R_b$ is very weak (see Fig.
10(b)).

\begin{figure}[htb]
\begin{center}
\scalebox{0.4}{
\includegraphics[width=20.0cm,clip]{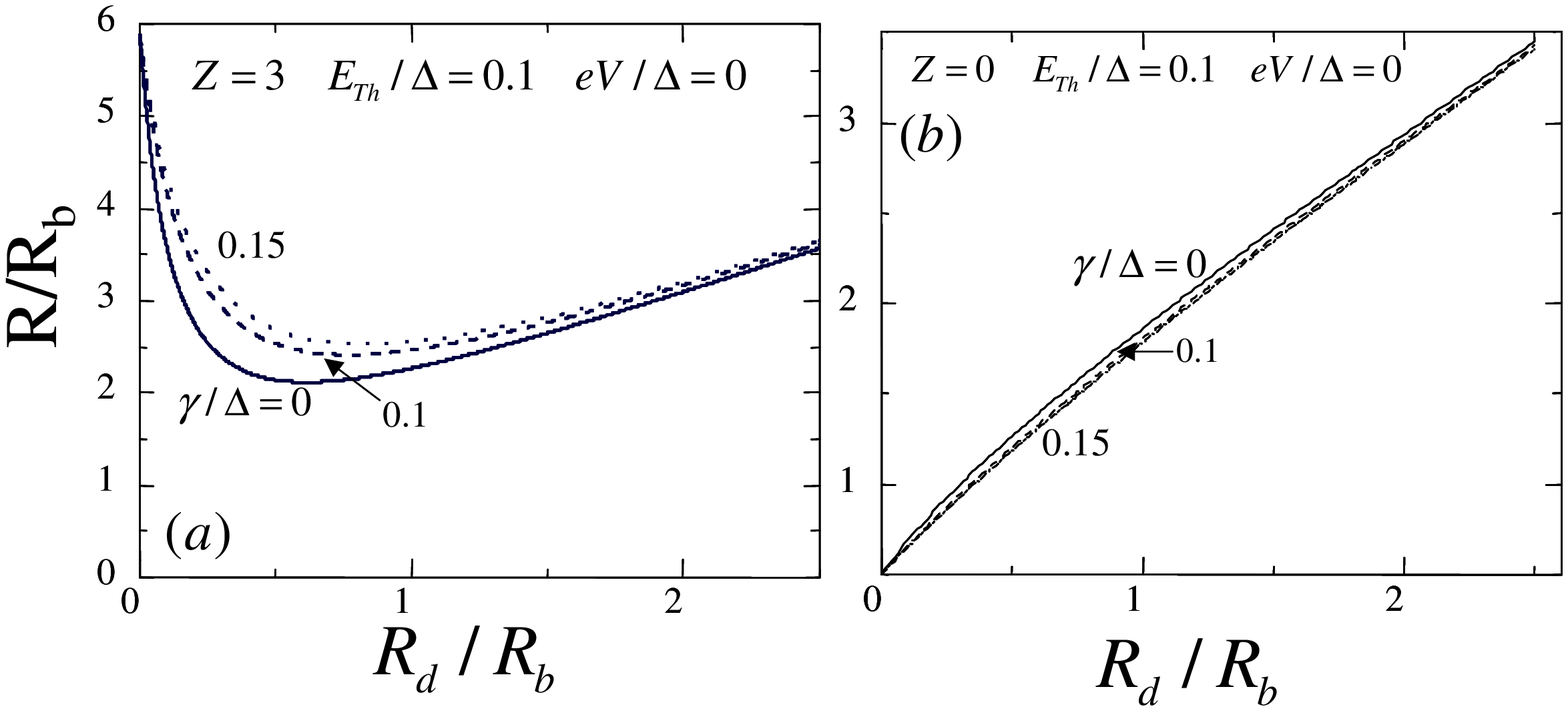}}
\end{center}
\par
\label{f10}
\caption{ Normalized zero voltage resistance as a function of $R_{d}/R_{b}$.}
\end{figure}

\begin{figure}[htb]
\begin{center}
\scalebox{0.4}{
\includegraphics[width=15.0cm,clip]{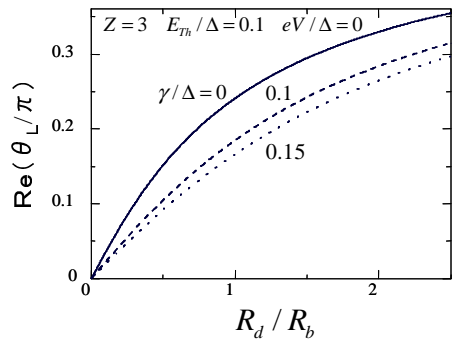}}
\end{center}
\par
\label{f11}
\caption{ Real part of $\protect\theta _{L}$ at zero energy as a function of
$R_{d}/R_{b}$.}
\end{figure}

\subsection{Tunneling conductance for $d$-wave junctions}

Below we discuss the results of calculations for the $d$-wave case. Fig. \ref%
{f12} shows the normalized conductance for $Z=10$, $R_{d}/R_{b}=1$, $%
E_{Th}/\Delta=0.01$, and $\alpha/\pi=0$ where $\alpha$ denotes the the
misorientation angle between the normal to the interface and the crystal
axis of $d$-wave superconductors. In this case, MARS are not formed at the
interface of the $d$-wave superconductor. The origin of the ZBCP is due to
the proximity effect in the DN region and the height of the ZBCP is
suppressed with increasing $\gamma$ similar to the case of the $s$-wave
junctions.

With the increase of the magnitude of $\alpha$ the MARS are formed at the
interface. The MARS contribute to the charge transport across the junction
and leads to the formation of the ZBCP. As is seen in Fig. 13, the ZBCP does
not depend on $\gamma$ for $Z=10$, $R_{d}/R_{b}=1$, $E_{Th}/\Delta=0.01$ and
$\alpha/\pi=0.125$. The similar result is obtained for different angle $%
\alpha/\pi=0.25$. The reason is that MARS reduce the proximity effect in DN,
therefore the influence of magnetic impurity scattering on the $\sigma_{T}$
becomes less important. In the extreme case, $\alpha=0.25\pi$, the proximity
effect is completely absent by the symmetry of the pair potential and $%
\sigma_{T}$ is completely independent of $\gamma$.

\begin{figure}[htb]
\begin{center}
\scalebox{0.4}{
\includegraphics[width=20.0cm,clip]{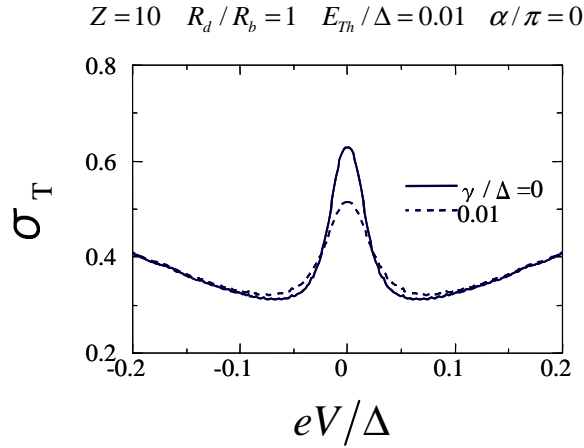}}
\end{center}
\par
\caption{ Normalized conductance in a $d$-wave junction for $Z=10$, $%
R_{d}/R_{b}=1$, $E_{Th}/\Delta=0.01$, and $\protect\alpha/\protect\pi=0$.}
\label{f12}
\end{figure}

\begin{figure}[htb]
\begin{center}
\scalebox{0.4}{
\includegraphics[width=20.0cm,clip]{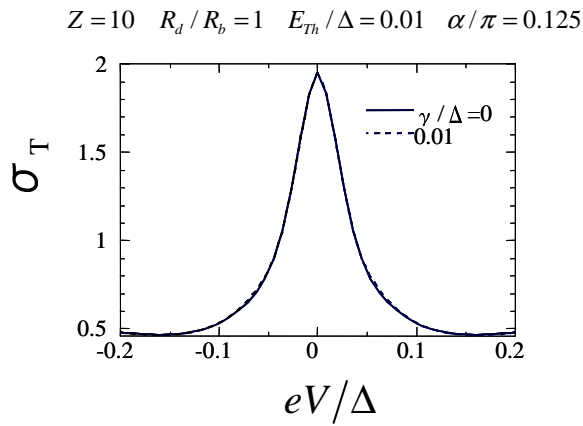}}
\end{center}
\par
\label{f13}
\caption{ Normalized conductance in a $d$-wave junction for $Z=10$, $%
R_{d}/R_{b}=1$, $E_{Th}/\Delta=0.01$, and $\protect\alpha/\protect\pi=0.125$.
}
\end{figure}

\section{Conclusions}

We have performed a detailed theoretical study of the conductance of
diffusive normal metal / $s$- and $d$-wave superconductor junctions in the
presence of magnetic impurities. Below the main results obtained in this
paper are summarized.

1. For the $s$-wave junctions, the proximity effect is suppressed by the
magnetic impurity scattering within the energy range determined by the Thouless
energy in DN. In this range both the real and imaginary parts of the
proximity effect parameter, i.e., $\mathrm{Re}(\theta_{L})$ and $\mathrm{Im}
(\theta_{L})$ are reduced with the increase of the magnitude of $\gamma$ for
any transparency of the insulating barrier.

2. The magnitude of the normalized bias voltage dependent conductance $%
\sigma_{T}(eV)$ in the low transparent $s$-wave junctions is suppressed by
the magnetic impurity scattering. On the other hand, for high transparent $s$%
-wave junctions, $\sigma_{T}(eV)$ can be enhanced by the magnetic impurity
scattering. 

3. In the $d$-wave junctions, the zero bias conductance peak formed for low
transparent barriers is suppressed by the magnetic impurity scattering only
for $\alpha \sim 0$. For other misorientation angles the conductance is not
sensitive to the magnetic impurity scattering in a diffusive normal metal.

In the present paper, we have discussed the case where magnetic impurities
are located in DN. These results can be also applied to the situation when
the junction is in a weak magnetic field $H$. If the field direction is
parallel to the junction plane, the pair-breaking rate is given by $\gamma
=e^{2}w^{2}DH^{2}/6$, where $w$ is the transverse size of the DN\cite%
{Belzig1}. Assuming $w=10^{-5}m$, $D=10^{-2}m^{2}/s$, $\Delta =10^{-3}eV$,
and $H=10^{-4}\sim 10^{-2}T$, we estimate the pair-breaking rate $\gamma
/\Delta =10^{-3}\sim 10$. This range of $\gamma $ corresponds to the
parameters chosen in the present paper. The suppression of the ZBCP and ZBCD
by the magnetic field was actually observed in several experiments\cite%
{Giazotto,Kastalsky,Bakker,Xiong,Magnee,Poirier}. The results of the present
paper may serve as a guide to study the charge transport in the junctions
with magnetic impurities or under applied magnetic field.

It is also an interesting problem to study the influence of the magnetic
impurity scattering on diffusive normal metal / triplet superconductor
junctions where anomalous proximity effect is expected \cite{pwave}. An
application of the present theory to the S/N/S junctions with unconventional
superconductors also requires separate investigation.


%
The authors appreciate useful and fruitful discussions with Yu. Nazarov and
H. Itoh. This work was supported by the Core Research for Evolutional
Science and Technology (CREST) of the Japan Science and Technology
Corporation (JST) and a Grant-in-Aid for the 21st Century  COE 
"Frontiers of Computational Science". The computational aspect of this work has been performed
at the facilities of the Supercomputer Center, Institute for Solid State
Physics, University of Tokyo and the Computer Center. 
%


\end{document}